# TRUST BASED PARTICIPANT DRIVEN PRIVACY CONTROL IN PARTICIPATORY SENSING

Ramaprasada R. Kalidindi[1], KVSVN Raju[2] V. Valli Kumari[3] and C.S. Reddy[4]

[1]Department of Computer Science and Engineering,
S.R.K.R. Engineering College, Bhimavaram-534204, India.
`rrkalidindi@computer.org`
[2,3,4]Department of Computer Science and Systems Engineering,
College of Engineering (A), Andhra University, Visakhapatnam-533003, India.
[2]`kvsvn.raju@gmail.com`, [3]`vallikumari@ieee.org`, [4]`csatyanand@gmail.com`

## ABSTRACT

*Widespread use of sensors and multisensory personal devices generate a lot of personal information. Sharing this information with others could help in various ways. However, this information may be misused when shared with all. Sharing of information between trusted parties overcomes this problem. This paper describes a model to share information based on interactions and opinions to build trust among peers. It also considers institutional and other controls, which influence the behaviour of the peers. The trust and control build confidence. The computed confidence bespeaks whether to reveal information or not thereby increasing trusted cooperation among peers.*

## KEYWORDS

*Pervasive computing, Participatory sensing, Confidence, Trust, Privacy control, Ubiquitous computing*

## 1. INTRODUCTION

Participatory sensing applications have attracted significant interest in recent years [1-5]. In this, participatory entities are ranging from networked sensors to multisensory personal devices. These sensors will generate high granular personal data. Protecting this personal data is a concern in these applications. Present applications were focused on centralized systems that provide participant privacy, protect data from misuse and check the integrity of data [6]. However, the cost of these systems increases exponentially with increase in system size. When these applications were provided in a distributed system, the control over participant's data should be with the participant. This participant driven privacy control will enhance people's participation in participatory sensing. In this, the participant has to decide which data is to be shared with whom [7].

To accomplish this, trusted peers have to be identified to exchange data. Generally trusted users tend to behave positively, where as distrusted users tend to behave negatively [8]. Otherwise, controls are of the need to protect the privacy of the participants. Government can provide these controls by enacting laws. However, with evolving cyber laws to create controls in digital domain and their delayed enactment raises the importance of trust in digital communities. With rapid advancements in networking and their penetration into society, there is a need to represent social trust in the form of digital trust [9]. This paper attempts to digitize social trust to exchange personal information by identifying confidence in peer taking interactions between them into consideration. Depending on this confidence personal data can be disclosed.

In this paper, section 2 describes related work; section 3 gives background of trust and describes factors that influence confidence, section 4 describes the proposed model to calculate confidence in peers so that information can be shared with those peers that are having higher confidence; and section 5 concludes the paper and suggests possible future directions.

## 2. RELATED WORK

Molm et al. [10] demonstrated the importance of reciprocal exchange in trust establishment over negotiated exchange in social environment. In this reciprocal exchange the risk and uncertainty of exchange provides an opportunity to demonstrate the partner's trustworthiness, which in turn helps in the development of trust in the partner.

Wagealla et al. [7] identified the need for privacy control at information owner's end and presented a trust based privacy control model for context aware systems. Braynov [11] described learning trust through interactions by observing the frequency of honoring and abusing trust. Wang and Varadharajan [12] presented a trust evaluation model to measure credibility of recommendations given by peers to obtain accurate trust values.

Yan and Holtmanns [13] summarized perceptions of trust and their influencing factors and characteristics from the literature. Wang et al. [14] presented a trust evaluation method to avoid dishonest sellers in peer-to-peer environment and indicating the risk for future transactions.

Kalidindi et al. [15] described a model for participant driven privacy control in participatory sensing for better participation. In this model, users can exchange different types of data to different users depending on the authorization levels. Ruizhong et al. [16] presented a dynamic trust model based on perceived risk considering user's subjective perception factors.

## 3. BACKGROUND

In society or in a networked environment, people may not be familiar initially, but overtime they are acquainted. When they are familiar, they form relationship and these relations are essential for ready cooperation when there is a need. These relations will improve with reciprocal interactions or gestures rather than negotiating relationship or with some binding agreement. These reciprocal interactions are evaluated perpetually. When there were no interactions, people tend to collect opinions (about one's reputation) from other people who are having direct interactions [17]. People establish relations independently in a distributed environment, where as in a centralized environment the relationship passes through a central authority. For example, if Alice wants to interact with Bob in a distributed system (Fig.1.a), Alice has to trust Bob directly. Alice should establish the trustor and trustee relationship by taking her past interactions with Bob and opinions from neighbours about their relationship with Bob. If Alice wants to interact with Bob in a centralised system (Fig.1.b), Alice has to trust an integrator Carl, and Carl has to trust Bob. Carl manages relationships, Alice and Bob need not worry about the relationship once if they are part of the system. These systems are also implemented in computing environment. In the computing environment, peer-to-peer systems are attracting significant interest because of their decentralization, scalability and organic growth with cooperation among users.

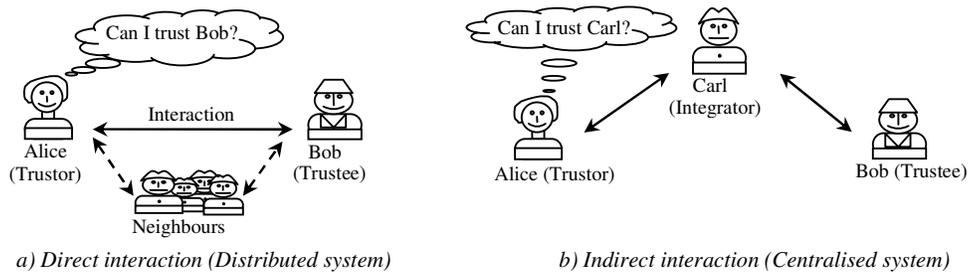

*a) Direct interaction (Distributed system)*                    *b) Indirect interaction (Centralised system)*

Figure 1. Alice interacting with Bob

In any environment, trust plays a major role in establishing relationship to facilitate cooperation among people. Relationship builds up when trust in other person increases. For this, the other person has to show trustworthiness in a reciprocal interaction [10, 13]. Trust is the estimate of expectation that the other person will perform as per one's need. This does not guarantee that the other person will perform as per the need. The entity that influences other person to perform is control and it provides assurance. This assurance comes from the knowledge of an incentive/punishment structure that encourages benign behavior or discourages misbehavior. That is, the trustor's trust on trustee and the control on trustee will influence the trustor's confidence on trustee [18]. The trustor's confidence expects trustworthy behavior from trustee. However, the trustor does not have control over the trustee's behavior. The trustee will act according to his trustworthiness. Trustworthiness is the internal intention of trustee, which translates to trustworthy behavior. Though there is no guarantee for trustworthy behavior, certain factors will influence this behavior. These factors will control the misbehavior or encourages benign behavior, for example, a driver is driving properly because of traffic rules set by law; an employee will follow timings fearing that employer may take action as prescribed in job agreement; etc.

As we know, trust and control are mutually exclusive. If the trustor has more control over the trustee, there is less need for the trust on trustee. Trust is a deficiency of control and is a desire to progress despite the inability to control. Symmetrically, control is a deficiency of trust. In other words, trust relates to the state of being dependent, while control relates to the ability to keep someone in a dependent position [11, 18, 19]. These trust and control build confidence [20], which is the belief of a person whether to rely on others or not. The control parameters that influence the trustworthy behavior of trustee will also contribute to confidence. Repetitive trustworthy behavior of trustee will increase the trustor's confidence on trustee. In the absence of interactions, opinions or control structures, the relationship would be a case of faith or gambling [21]. Fig.2 shows factors influencing the trustor's confidence on trustee.

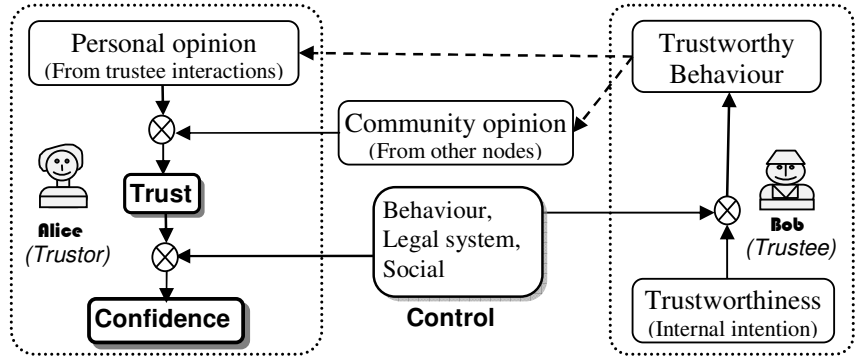

Figure 2. Factors influencing the confidence

# 4. PROPOSED MODEL

In this model, we describe a network of computer systems placed in each flat of a gated residential community. The computer system in each flat manages networked sensors installed in it. These computer systems interact with each other as nodes in the network. A node forms personal opinion based on positive interactions. Trust evaluation considers personal opinion and opinions from other nodes. Finally, trust and other controlling parameters that influence the trustworthy behavior of the trustee build confidence on another node. We assume: 1) Network is robust and free from failures and propagation delays are negligible. 2) Nodes are honest (not correct always). 3) If a request comes, the node may or may not respond.

Let there be $n$ nodes ($N_1 - N_n$) in a network as shown in Fig.3. The solid lines are interactions that involve information exchange. The dotted lines are requests for opinions about others. The node $N_i$ (trustor) is evaluating trust with $N_j$ (trustee) to share data.

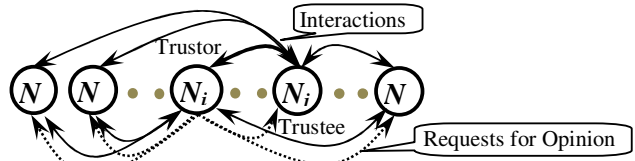

Figure 3. Nodes' interactions for establishing trust

## 4.1. Interactions

An interaction can be defined as the communication between nodes to exchange information. The nodes may also request for opinions, which are not interactions. The response to the request is an interaction. The interactions may be positive or negative depending on scores assigned by requesting node. If $N_i$ has a total of $t_{ij}$ requests to $N_j$ and among these $t_{ij}^p$ are positive and $t_{ij}^n$ are negative responses depending on various factors represented by interaction properties [14, 16]. The status (positive or negative) of the interaction is determined by calculating numerical scores for each property. Depending on the status of overall interactions, personal opinion (*ib.* section 4.2) is calculated.

Let $P = \{p_1, p_2, \ldots\ldots, p_m\}$ be the set of $m$ properties for an interaction and $I_k$ is the score for $k^{th}$ property. The $I$ is the aggregated weighted score which determines whether an interactions is positive or not. We consider the following five properties for an interaction to determine whether it is positive or not.

### 4.1.1. Response time (p₁)

In many social situations when we ask for help from a neighbour, his prompt response (especially in the case of emergencies) for our request builds relationship. The response time is the elapsed time between request made by the requesting node and to the response from the responding node. This reflects the importance the responding node is giving to request. Less response time gives more weightage for the interaction. Response time variations resemble inverse Gompertz curve [22] as shown in Fig. 4, where decay is slowest at the start and end of a period and is given as:

$$I_1 = 1 - e^{-be^{-ct}} \qquad (1)$$

Where, $I_1$ is the score; $b$ and $c$ are constants and $t$ is the time in hours. The node cannot wait infinitely for the response to a request. It will wait for certain amount of time to collect the responses from various nodes and inference is taken. The $b$ determines waiting time threshold and $c$ determines window in which the response is accepted with reduced weightage.

The response time score $I_1$ is calculated by selecting $b$ to individual node's perspective that how much time it has to wait for the response? After selecting $b$ value, $c$ is selected as 0.5 such that there will not be any relevance for the response if it reaches after waiting time. Fig.4 shows the forms of score variation for different $b$ values when $c$=0.5.

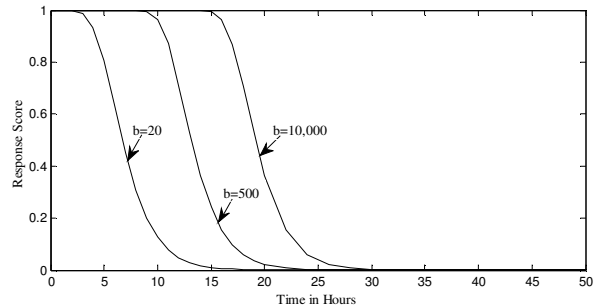

Figure 4. Response scores for different values of $b$ when $c=0.5$

### 4.1.2. Time gap (p₂)

Relationship between nodes develops gradually or fails to develop. Beneficial acts of requesting node prompt reciprocal benefit from responding node, which result in frequent interactions. Continuous interactions result in better relationship where as discrete interactions show negligible relationship. The relation takes the form of a series of sequentially contingent acts; for example, our neighbour cares for our house while we are gone for a tour, we will bring gifts, he invites us to lunch when we return from tour, and so forth.

Continuous interactions can be identified by taking the inter interaction time gap. The time gap is the interval between present interaction and the immediate previous interaction of the requesting or responding node. Lesser time gap indicates extending relationship, higher time gap indicates discrete relationship. Indirectly this property gives higher weightage for the interaction from a node, which is maintaining continuous relationship. Time gap variations also resemble inverse Gompertz curve [22] as shown in Fig. 5, and is given as:

$$I_2 = 1 - e^{-be^{-ct}} \qquad (2)$$

Where, $I_2$ is the score; $b$ and $c$ are constants and $t$ is the time in months.  The $b$ determines maximum time gap threshold between interactions in continuous interactions; $c$ determines window in which present interaction is related to immediate previous interaction for determination of continuous interactions, which will have higher weightage.

The time gap score  $I_2$ is calculated by selecting $c$ value to individual node's perspective on whether the interaction is occasional or frequent and b value is taken as 10 because we need not wait to determine the interaction as a continuous one or not.  As the inter interaction time gap increases the relationship ceases to be a continuous relationship.  The Fig.4 shows the forms of score variation for different $c$ values when $b$=10.

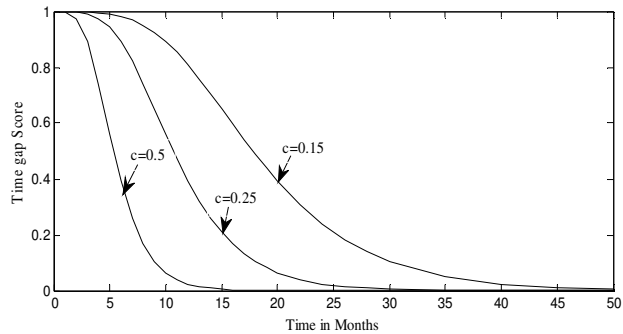

Figure 5. Time gap score for different values $c$ when $b=10$

### 4.1.3. Familiarity (p₃)

Usually long lasting relationships in society will have more weightage than recent relationships.  Over time, as the interactions between requesting node and responding node increase the familiarity increases.

Familiarity depends on the acquaintance period between the nodes.  As this period increases familiarity increases and this increase resembles Gompertz curve [22] as shown in Fig.6, where growth is slowest at the start and end of a period.  The score for this property is given as:

$$I_3 = e^{-be^{-ct}} \qquad\qquad (3)$$

Where, $I_3$ is the score for the familiarity; $b$ and $c$ are constants and $t$ is the time in years.  The curve forms for various $c$ values and $b$=10 are shown in Fig.6.  The selection of $b$ and $c$ represents the node's perspective about the familiarity, where low $c$ value takes more time and high $c$ value takes less time to get full familiarity.  The $b$ value is taken as 10.  As time passes familiarity increases, thereby we need not wait for starting familiarization.

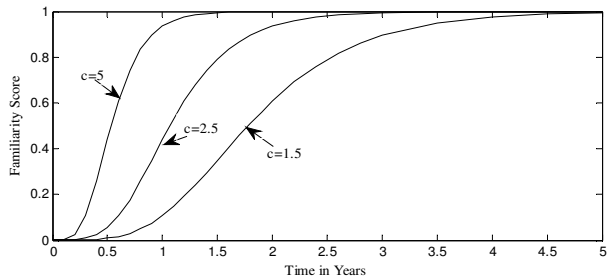

Figure 6. Familiarity score for different values $c$. when $b=10$

### 4.1.4. Reciprocity (p₄)

In our daily life, reciprocal acts of benefit, viz. offering help, advice, approval etc., form social exchange [10]. This reciprocity is an indicator of stability in social relations [23]. In building relationships, people observe how far the other person is reciprocating for their gestures. Reciprocal exchanges build relationships. Lack of reciprocity ruins relationships.

In this paper, reciprocity is considered as the extent of information (viz. context or personal) that responding node is sharing with requesting node or it can be authorization level given by responding node to requesting node to access the information. The nodes divide their privacy levels according to their convenience, viz. 3, 5, 10 levels or as percentage. The highest level is free access (zero privacy) and lowest level is no access (full privacy). One of these levels is allowed as a reciprocal gesture by the responding node to the requesting node. Since the requesting node can access information, we assume that it knows the number of levels (and also max., min. levels) and its privilege level. To get normalized score for different nodes the reciprocity score $I_4$ is given as:

$$I_4 = \frac{r - r_{\min}}{r_{\max} - r_{\min}} \qquad (4)$$

Where, $r$ represents the privilege/authorization level given by the responding node to the requesting node; $r_{\max}$ is the maximum level (free access) at the responding node; $r_{\min}$ is the minimum level (no access) at the responding node and $r_{\min} \leq r \leq r_{\max} \in [0,1]$.

### 4.1.5. Relevance (p₅)

When we approach for help to our neighbors, if the response from our neighbors is not to our expectations it will disappoint us. Some may give irrelevant gestures, which may not be useful to us. Relevance property reflects the relevance of the response for the request made by requesting node.

Relevance is a subjective parameter. To determine the score for these types of properties a grading level $F_k$ for each property $P_k$ is assigned as: $F_k = G_k(p_k)$. Where $G_k$ is the grading function of property $p_k$, which converts property value into the corresponding grading level.

Score intervals $I_k$ of each property are calculated as: $I_k = S_k(F_k)$. Where $S_k$ is the score function that maps grading level. For relevance score the grading levels are mapped to five score intervals [0, 0.25, 0.5, 0.75, 1]. Table 1 gives scores for corresponding grading levels.

Table 1. Grading levels and Scores for Relevance

| Grading levels ($F_k$) | Score ( $I_k$ ) |
|---|---|
| Not at all relevant | *0.00* |
| May not be relevant | *0.25* |
| Can't say | *0.50* |
| To some extent relevant | *0.75* |
| Fully relevant | *1.00* |

Usually not all properties of interaction are equally important; some properties (viz. reciprocity and familiarity) have more influence than others do.  So user can give relative importance using weight $w_k^p$ to each property, such that $w_k^p \in [0,1]$, $\sum_{k=1}^{m} w_k^p = 1$.

Table 2 gives weightages for each interaction property.  A node can set these weightages according to its perception about an interaction.

Table 2.  Interaction Property Weightages

| Interaction Property ($p_k$) | $w_k^p$ |
|---|---|
| Response time ($p_1$) | 0.2 |
| Time gap ($p_2$) | 0.1 |
| Familiarity ($p_3$) | 0.3 |
| Reciprocity ($p_4$) | 0.3 |
| Relevance ($p_5$) | 0.1 |

The sum of weighted scores $I$ is given as:

$$I = \sum_{k=1}^{m} \left( w_k^p \times I_k \right) \qquad (5)$$

The interaction is positive if the value of $I$ crosses a threshold.  For example, a user may consider taking data as shown in Table 3, where $b$ and $c$ are constants and $t$ is the time factor for each property in equations (1-3).

Table 3.  Sample data for Interaction properties

| Property | b | c | t |
|---|---|---|---|
| Response time | 500 | 0.5 | 10 Hrs. |
| Time gap | 10 | 0.25 | 5 Months |
| Familiarity | 10 | 2.5 | 1Year |
| Reciprocity | 9th level on 10 level scale | | |
| Relevance | Fully relevant | | |
| Threshold for $I$ value is 0.5 | | | |

The value of $I$ for the above data is 0.7894 by equations (1-5).  Since it is greater than the threshold, this is a positive interaction.  The node, depending on its perspective about positive interaction chooses the threshold.  Low threshold indicates optimistic view of other node's interactions and high threshold indicates pessimistic view.

## 4.2. Opinion

As per the Oxford dictionary definition, the *opinion* is "a view or judgment about a particular thing, which is not necessarily based on fact or knowledge" [24].  In social interactions, our relationship will vary with stranger to the acquaintance.  People trust acquainted persons depending on the interactions with them and stranger too trusted to some extent.  People will form personal opinion based on interactions they have and considers the opinion of the acquainted persons.  If a person is having a minimum number of interactions to judge, then there is no need for taking the other's opinion. These opinions will have weightage depending on the trust that a person is having on the acquainted person. Mostly in this type of situations, the opinions of trusted persons will influence the decision making process. The opinion may be positive or negative. Here we consider two opinions for identifying the trust on the trustee.

### 4.2.1. Personal opinion

Personal opinion is the value given by a node depending upon the responses it is having from responding node. The node $N_i$'s *personal opinion, $Op_{ij}^p$* on $N_j$ is the ratio of effective interactions to total requests (It is the total of positive, negative interactions and no responses) and $Op_{ij}^p \in [-1,1]$. The effective interactions are the difference between positive and negative interactions. The self opinion value is one (i.e., $Op_{ii}^p = 1$). All interactions to the node itself are positive interactions (i.e., $t_{ii} = t_{ii}^p$).

$$Op_{ij}^p = \frac{t_{ij}^p - t_{ij}^n}{t_{ij}} \qquad (6)$$

After determining whether the interaction is positive or negative using equation (5) the personal opinion is calculated using equation (6). If $Op_{ij}^p < 0$, $N_i$ is having negative opinion on $N_j$, for $Op_{ij}^p > 0$ $N_i$ is having positive opinion on $N_j$. For $Op_{ij}^p = 0$, $N_i$ is not having any opinion. Each node gives a weightage to a node depending on its opinion about that node. The weightage $w_{ij}^n$ given to a node $N_j$, at $N_i$ is equal to the value of the personal opinion if it is positive and zero for other values.

$$w_{ij}^n = \begin{cases} Op_{ij}^p & \text{if } t_{ij}^p > t_{ij}^n \\ 0 & \text{if } t_{ij}^p \le t_{ij}^n \end{cases} \qquad (7)$$

Each node will maintain an opinion table, as in Table 4, which contain its opinion on other nodes, their weightage and total number of interactions with those nodes in the network.

Table 4. Opinion Table at Node $N_i$

| Node | Node weightage | Opinion | Interactions | | |
|------|----------------|---------|--------------|---|---|
| | | | Positive | Negative | Total |
| $N_1$ | $w_{i1}^n$ | $Op_{i1}^p$ | $t_{i1}^p$ | $t_{i1}^n$ | $t_{i1}$ |
| . | . | . | . | . | . |
| . | . | . | . | . | . |
| $N_n$ | $w_{in}^n$ | $Op_{in}^p$ | $t_{in}^p$ | $t_{in}^n$ | $t_{in}$ |

For every request made by the node, it updates the total number of interactions. The node will wait for response for a prescribed time, after this time is over it sends another request. The node also updates opinion and node weightage with every interaction by calculating these values using equations (6) and (7) respectively as shown in Fig.7.

### 4.2.2. Community opinion

Community opinion is the public opinion collected from other nodes in the network. When a node $N_i$ wants to calculate the opinion on node, $N_j$ it sends a request to all other nodes. These nodes will respond to this request by sending the opinion on $N_j$, which is available in their respective opinion tables. The received opinion will have importance depending on responding node's weightage.

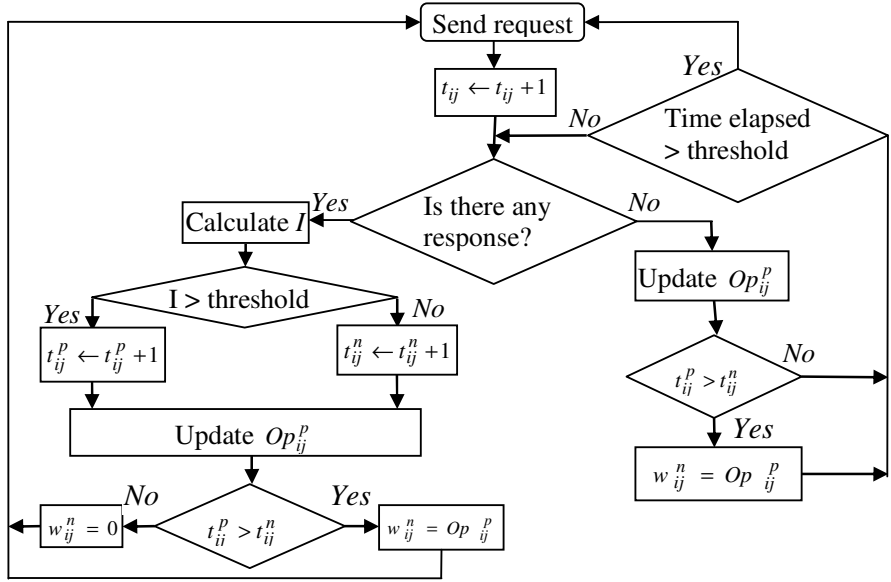

Figure 7. Flow chart for updating opinion table (*ib.* Table 4)

A node that is having more interactions with other node can give better opinion about that node. As the number of interactions *(i)* increase, the effect of new *(i+1)* interaction on opinion is minimal. The inferred opinion from other nodes, *community opinion*, $Op_{ij}^c$ is given as:

$$Op_{ij}^c = \frac{\sum_{x=1}^{n} \left( w_{ix}^n \times Op_{xj}^p \right)}{n} \qquad (8)$$

Where $Op_{xj}^p$ is the opinion given by node $N_x$ to node $N_i$; $w_{ix}^n$ is the weightage of the node $N_x$ at the node $N_i$; and *n* is the number of nodes responding to the request for opinion.

## 4.3. Trust

Trust may depend on context, where a person trusts another in a particular situation and it can also be independent. The independent trust which is also called as reliability trust gives an idea of person's general reliability. Jøsang et al. [25] defined reliability trust as the subjective probability by which trustor expects trustee to perform a given action on which trustor's welfare depends. Here context independent trust is considered. In order to form trust, the trustor must have enough knowledge about the trustee. Such knowledge comes from interactions with trustee and publicly established trustee's reputation in community. After calculating the community opinion using equation (8) the trust is obtained as:

$$Tr_{ij} = \begin{cases} Op_{ij}^p & \text{if } t_{ij} \geq t_{\min}^i \\ Op_{ij}^p \otimes Op_{ij}^c & \text{if } t_{ij} < t_{\min}^i \end{cases} \qquad (9)$$

Where the trust $Tr_{ij}$ is node $N_i$'s trust on $N_j$, such that $Tr_{ij} \in [-1,1]$ and $t_{\min}^i$ is the minimum number of interactions needed by node $N_i$ to trust another node on its own without going for community opinion. This will depend upon the node's judgment capability and it varies from one node to another node. The operator $\otimes$ obtains trust value $Tr_{ij}$, depending upon personal opinion $Op_{ij}^p$ and community opinion $Op_{ij}^c$ values as shown in Table 5.

Table 5. Operations of $\otimes$ operator

| $Op_{ij}^p$ | $Op_{ij}^c$ | $Tr_{ij}$ |
|---|---|---|
| 0 | - | $Op_{ij}^c$ |
| - | 0 | $Op_{ij}^p$ |
| < 0 | < 0 | |
| > 0 | > 0 | |
| > 0 | < 0 | $\begin{cases} Op_{ij}^p & \text{if } \left\|Op_{ij}^p\right\| > \left\|Op_{ij}^c\right\| \\ Op_{ij}^c & \text{if } \left\|Op_{ij}^p\right\| < \left\|Op_{ij}^c\right\| \\ 0 & \text{if } Op_{ij}^p = Op_{ij}^c \end{cases}$ |
| < 0 | > 0 | |

In some cases individual knowledge (personal opinion) and public reputation (community opinion) could be conflicting and these are opposite and equal, i.e., $Op_{ij}^p = -Op_{ij}^c$, this shows the node's mischievous behavior which is a different from zero trust without transactions.

## 4.4. Control

In society, certain factors influence people to be trustworthy. These factors are behaviour that gives person orderliness in daily life; legal system that compels people to follow common rules; social norms that establish common culture and values; etc. The control parameters that influence the trustworthy behavior of trustee will also contribute to confidence.

Let $C$ be the set of $q$ parameters $C = \{c_1, c_2, \ldots, c_q\}$ that influence the trustworthy behavior of a node and $Cf_{ip}^p$ is the node $N_i$'s confidence in these parameters to influence trustee. The relative importance of these parameters is given by node $N_i$ as control weight $w_{ip}^c$ such that $\sum_{p=1}^{q} w_{ip}^c = 1$.

The user may consider control parameters and his confidence on these parameters and their relative weights as shown in Table 6.

Table 6. Control parameters.

| Parameters ($c_q$) | $Cf_{ip}^p$ | $w_{ip}^c$ |
|---|---|---|
| Behaviour ($c_1$) | 1.0 | 0.2 |
| Legal system ($c_2$) | 0.8 | 0.5 |
| Social ($c_3$) | 0.5 | 0.3 |

The $Cl_{ij}$ is the control on node $N_j$ by control parameters as perceived by $N_i$ is given as:

$$Cl_{ij} = \sum_{p=1}^{q} \left( w_{ip}^c \times Cf_{ip}^p \right) \qquad (10)$$

## 4.5. Confidence

Confidence is a belief that trustor can have faith in trustee. That is, trustor expects something to happen with certainty, and does not consider the possibility of anything going wrong. Confidence expects trustworthy behavior from trustee, whereas trust expects trustworthiness. Control gives some sort of assurance. This assurance is the expectation based on knowledge of a punishment/incentive structure to encourage good behaviour [13]. Both trust and control contribute in building confidence. The trustor node $N_i$'s confidence on trustee node $N_j$ is:

$$Cf_{ij}^n = Tr_{ij} + Cl_{ij} \qquad\qquad (11)$$

The control $Cl_{ij}$, which influences trustworthy behavior as shown in Fig.2 is obtained by considering the values given in Table 6 by equation (10). This control and trust $Tr_{ij}$ from equation (9) are added to obtain confidence $Cf_{ij}^n$ as in equation (11). After calculating confidence the node, $N_i$ can establish trustful relationship with $N_j$ to share information. The user may choose a threshold level. If confidence is greater than a threshold level, it can share information otherwise not. Higher threshold level allows data sharing with only highly trusted people (desire for more privacy) and lower level allows data sharing with more number of people (less bothered about privacy).

## 5. CONCLUSION AND FUTURE DIRECTIONS

In a residential community having hundreds of houses at a particular place, people may install various sensor networks in their flats for surveillance and other purposes and may share this data with others. In this type of urban sensing environment, there is a need for a proper trust management between the collaborative entities. People may demand control over what they share and what they do not. Privacy control at the source will enable willing and engaged participation of citizens to create urban infrastructure with reduced cost.

This work considered the problem of establishing confidence in neighbours in a sufficiently large residential community by collecting opinions from others. Depending on confidence levels, people can authorize others to access data. In many social situations, confidence in other person is considered for sharing information or for delegating work. Confidence evaluation under malicious behaviour of nodes; collusion between nodes to get authorization; and considering risk factor along with trust to exchange data are the areas for further study to have a robust trust management for participatory sensor networks.

## ACKNOWLEDGEMENTS

The authors would like to thank DENSE research group members for their valuable inputs in their discussions.

**Authors**

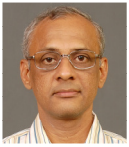 **Ramaprasada Raju Kalidindi** received the B.E. degree in Electronics and Communication Engineering from Andhra University, India and M.Tech. degree in Computer Science and Technology from IIT, Roorkee, India.  He is a member of ACM, IEEE Computer and Communication societies.  His research interests include Security and Privacy in Sensor Networks and Trusted Computing.

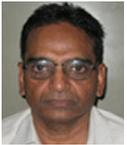 **KVSVN Raju** received the B.E. degree in Electrical Engineering from Government College of Engg., Kakinada, India and M.E. degree in Control Systems from Andhra University, India and obtained the PhD degree in Computer Science and Technology from IIT, Kharagpur, India.  His research interests include Data Engineering, Security Engineering and Software Engineering.

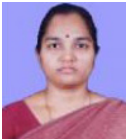 **V. Valli Kumari** received the B.E. degree in Electronics and Communication Engineering and M.Tech. degree in Computer Science and Technology and obtained the PhD degree in Computer Science and Technology from Andhra University, India.  Her research interests include Security and Privacy issues in Data Engineering, Network Security and E-Commerce.  She is a member of IEEE and ACM and is a fellow of IETE.

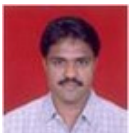 **C. Satyananda Reddy** received his M.Tech. degree in Computer Science (Software Engineering.) from JNTU, Hyderabad, India and obtained PhD degree in Computer Science and Technology from Andhra University, India.  His research interests include Software Engineering and Security Engineering.